\RequirePackage{fix-cm}
\documentclass[smallextended]{svjour3}       
\smartqed  
\usepackage{graphicx}
\begin{document}

\title{Quantum Information Transmission
}


\author{Lei Wang \and Jie-Hui Huang \and Jonathan P. Dowling \and Shi-Yao Zhu
}


\institute{Lei Wang  \at
Beijing Computational Science Research Center, Beijing 100084, China \\
College of Physics, Jilin University, Changchun 130012, China
 \\
              Tel.: +86-010-82687088\\
              Fax: +86-010-82687002\\
              \email{wang$\_$lei98@163.com}
           \and
           Jie-Hui Huang \at
              Beijing Computational Science Research Center, Beijing 100084, China
              \and
Jonathan P. Dowling \at Beijing Computational Science Research
Center, Beijing 100084, China \\
Hearne Institute for Theoretical Physics and Department of Physics
and Astonomy, Louisiana State University, Baton Rouge, Louisiana
70803 USA \and Shi-Yao Zhu \at Beijing Computational Science
Research Center, Beijing 100084, China}

\date{Received: date / Accepted: date}

\maketitle

\begin{abstract}
We present a scheme of quantum information transmission, which
transmits the quantum information contained in a single qubit via
the quantum correlation shared by two parties (a two-qubit channel),
whose quantum discord is non-zero. We demonstrate that the quantum
correlation, which may have no entanglement, is sufficient to
transmit the information of a quantum state. When the correlation
matrix of the two-qubit channel is of full rank (rank three), the
information of the qubit in either a mixed state or a pure state can
be transmitted. The quantum discord of a channel with rank larger
than or equal to three is always non-zero. Therefore, non-zero
quantum discord is also necessary for our quantum information
transmission protocol. 
\end{abstract}

\section{Introduction}

Quantum correlations have been demonstrated to be a resource for
computing [1,2], metrology [3-4], imaging [5-8], communication
[9,10] and steganography [11,12]. In each of these applications,
entanglement has been central to the advantage provided by quantum
systems. It is a question that has been discussed for quite a long
time is whether other kinds of quantum correlation (besides the
entanglement) can also be exploited in ways similar to entanglement.
One such quantum correlation is quantum discord [13-15]. States with
non-zero quantum discord are quantum correlated. While all entangled
states have non-zero quantum discord, there exist states of non-zero
quantum discord that possess no entanglement. In this paper we
proposed a new use of the quantum correlations: to transmit
information of an unknown qubit between two parties via a two-qubit
state with non-zero quantum discord shared by the two parties. The
scheme is similar to the quantum teleportation [16-19], and the
information of the qubit can be transmitted from one place to
another, if the correlation matrix of the two-qubit state (the
quantum channel) is of full rank (rank three). The quantum discord
of a state with full rank (rank three) is always non-zero. That is
to say, the quantum information transmission is possible using a
quantum channel with non-zero quantum discord. However, a quantum
channel with non-zero quantum discord might have no quantum
entanglement.

\section{Quantum information transmission}
Our set up is shown in Fig. 1, which is similar to the original
quantum teleportation scheme [16]. A source of quantum particles
generates a pair of qubits that have non-zero bipartite quantum
correlations. Alice and Bob share the two qubits. Alice (the sender)
makes a Bell measurement on her own qubit and an unknown single
qubit \emph{C} whose information will be transmitted to Bob (the
receiver), and sends Bob the measurement result via a classical
channel. The state Bob received will collapse to $\rho_B^\prime$. We
will derive the conditions under which the state $\rho_B^\prime$
contains complete information of the unknown state \emph{C}.
Non-classicality, which here is characterized by the rank of the
correlation matrix [15], plays an important role in the quantum
information transmission of the state of qubit \emph{C}.

\begin{figure}
\begin{center}
 \includegraphics[width=0.75\textwidth]{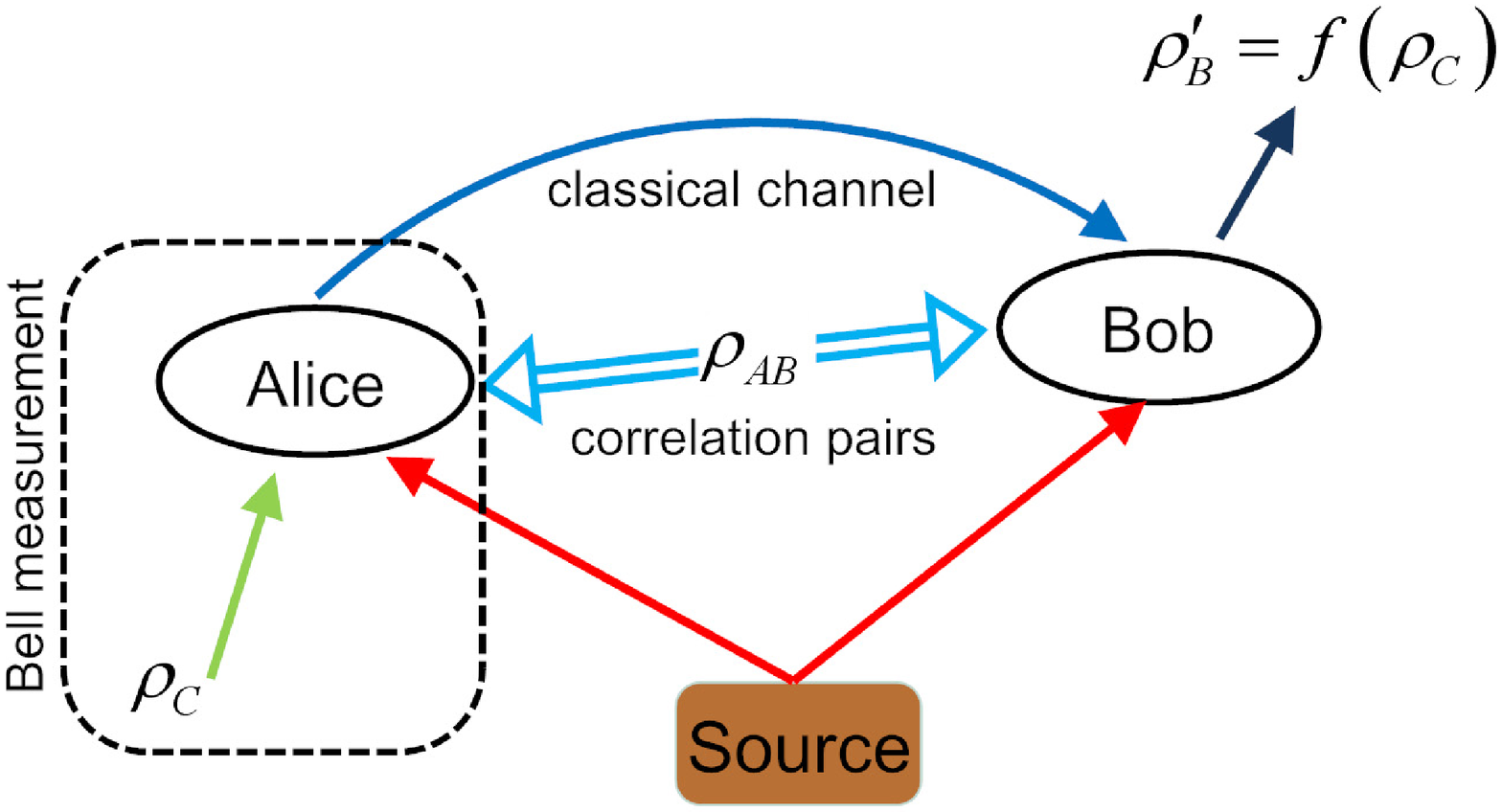}
 \end{center}
\caption{(Color online) Our scheme of the quantum information
transmission. The source sends two correlated qubits to Alice and
Bob. Once Alice performs a joint Bell measurement on her qubit
\emph{A} and the qubit \emph{C} (with state $\rho_\emph{C}$ that is
unknown to Alice or Bob), and sends Bob the result, the state of
Bob's qubit collapses to the mixed state $\rho_\emph{B}^\prime$.
Under some constraints, the state $\rho_\emph{B}^\prime$ contains
complete information of the unknown (possibly even mixed) state
\emph{C}.}
\label{fig:1}       
\end{figure}

In the general case, a two-qubit quantum channel system can be
written as,
\begin{eqnarray}\label{eq1}
\rho_{AB}=\frac{1}{4}\sum\limits_{i,j=0}^3r_{ij}\sigma_i^A\otimes\sigma_j^B\quad
\{-1\le{r_{ij}}\le1, r_{00}=1 \},
\end{eqnarray}
where $\sigma_0$ is the identity operator, $\sigma_i \{i=1,2,3\}$
are the three Pauli operators. The unknown single-qubit input state
can be written as:
\begin{eqnarray}\label{eq2}
\rho_{C}=\frac{1}{2}\sum\limits_{i=0}^3c_i\sigma_i^C\quad \{
-1\le{c_{i}}\le1, c_{0}=1\}.
\end{eqnarray}
where the information of qubit \emph{C} is contained in the three
real parameters, $c_i$  (\emph{i}=1,2,3). The expression
$\rho_C\otimes\rho_{AB}$ denotes the state of the entire system. Now
we consider the requirement for implementing the quantum information
transmission. We choose the Bell states as the measurement basis on
the \emph{AC} system. After the measurement, the state \emph{B} will
collapse to,
\begin{eqnarray}\label{eq3}
\rho_B^\prime=\frac{\langle\beta_{mn}|\rho_C\otimes\rho_{AB}|\beta_{mn}\rangle}{\rm{Tr}(\langle\beta_{mn}|\rho_C\otimes\rho_{AB}|\beta_{mn}\rangle)},
\end{eqnarray}
where $|\beta_{mn}\rangle$ $\{m,n=0,1\}$ is one of the Bell states,
and
$\rm{Tr}(\langle\beta_{mn}|\rho_C\otimes\rho_{AB}|\beta_{mn}\rangle)$
is the probability of the corresponding measurement outcome that is
used to renormalize the density matrix. According to Alice's four
different possible measurement outcomes, corresponding to the four
Bell states, there are four possible expressions for $\rho_B^\prime$
that can be written as,
\begin{eqnarray}\label{eq4}
\rho_B^\prime=\frac{1}{2}\left(
  \begin{array}{cc}
   1+s_3^{mn} & s_1^{mn}-is_2^{mn}  \\
    s_1^{mn}+is_2^{mn} & 1-s_3^{mn}  \\
  \end{array}
\right),
\end{eqnarray}
where
\begin{eqnarray}\label{eq5}
s_k^{mn}=\frac{r_{0k}+\sum\limits_{j=1}^3(-1)^{\phi_j}r_{jk}c_j}{1+\sum\limits_{j=1}^3(-1)^{\phi_j}r_{j0}c_j}\quad
\{k=1,2,3\},
\end{eqnarray}
where $\phi_1=m$, $\phi_2=m+n$, $\phi_3=n$. The denominator in Eq.
(\ref{eq5}) is equal to zero only when the state $\rho_{AB}$ is a
direct product state,  a completely classical and uncorrelated
states that will not be considered here. Now let us prove that the
state $\rho_B^\prime$ includes all the information needed to
reconstruct $\rho_C$. In general, the state $B^\prime$ has three
independent parameters, $s_k^{mn}$ $\{k=1,2,3\}$ in Eq. (\ref{eq4}),
for each of the four Bell states. Select $m=n=0$ ($s_k^{00}=s_k$) as
an example. The values of $m$ and $n$ are sent to Bob via the
classical channel. From Eq. (\ref{eq5}) we have
\begin{eqnarray}\label{eq6}
\sum\limits_{j=1}^3(-1)^{j+1}(r_{jk}-r_{j0}s_k)c_j=s_k-r_{0k} \quad
\{k=1,2,3\}.
\end{eqnarray}
From Eq. (\ref{eq6}) we construct three linear equations for $c_j$.
The information needed by Bob can be faithfully extracted when the
Eq. (\ref{eq6}) have a unique-solution, which means that the
coefficient matrix of the equations is of full rank. By comparing
the four matrices in Eq. (\ref{eq4}), we find that the four
coefficient matrices can be transformed into each other by
elementary matrix operations, which will not change the rank of a
matrix. This means that the unique-solution condition is the same
for each of the four density matrices in Eq. (\ref{eq4}). Thus we
only need to consider any one of the four. We define the coefficient
matrix for Eq. (\ref{eq6}) as \emph{T}
\{$T_{j,k}=(-1)^{j+1}(r_{jk}-r_{j0}s_k)$\}. By substituting $s_i$ in
Eq. (\ref{eq5}) into $T$, it is interesting to note that the
determinant of $T$ is connected to the determinant of the matrix
$R$, through the relation
$(1+r_{10}c_1-r_{20}c_2+r_{30}c_3)det(T)=-det(R)$, where the matrix
\emph{R} \{$R_{i,j}=r_{i-1,j-1}$\} is the correlation matrix defined
in Ref. [15]. The condition for unique solution for Eq. (\ref{eq6})
becomes that the \emph{R} matrix is of full rank. Therefore, in
order to implement the \emph{information} transmission of an
arbitrary qubit, full rank for the correlation matrix of the channel
\emph{AB} system is required.

\begin{figure}
\begin{center}
 \includegraphics[width=0.75\textwidth]{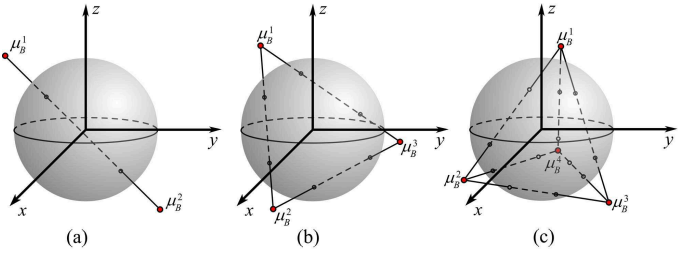}
 \end{center}
\caption{(Color online) The states in Bloch sphere. (a) (b) (c)
represent the collapsed states are located in line, plane and solid
space with rank=2,3,4, respectively.}
\label{fig:1}       
\end{figure}

In the following, we discuss the role of the rank of the correlation
matrix \emph{R} in the information transmission. The matrix $R$ can
be written as $R=UDW^T$ by singular value decomposition (SVD) [20],
where $U$ and $W$ are 4$\times$4 orthogonal matrices,
$D=\rm{diag}[\emph{d}_1\quad \emph{d}_2\quad \emph{d}_3\quad
\emph{d}_4]$, and the number of non-zero diagonal elements in
\emph{D} is the rank of \emph{R}. Comparing with Eq. (\ref{eq1}), we
have,
\begin{eqnarray}\label{eq7}
\rho_{AB}=\frac{1}{4}\sum\limits_{i=1}^4d_iA_i\otimes
B_i=\sum\limits_ip_i\mu_A^i\otimes\mu_B^i,
\end{eqnarray}
where $p_i=d_iU_{1i}W_{1i}$,
$\chi_i=\sum\limits_{j=1}^4\theta_{j,i}\sigma_{j-1}^\chi=\theta_{1,i}\sum\limits_{j=0}^3\alpha_j^i\sigma_j^\chi$
$\{\chi=A(or B)$ and $\theta=U(or W)\}$, and
$\alpha_j^i=\frac{\theta_{j+1,i}}{\theta_{1,i}}$. Note
$\sum_ip_i=1$, and the individual $p_i$ can be negative.
$\mu_\chi^i=\frac{1}{2}\sum\limits_{j=0}^3\alpha_j^i\sigma_j^\chi$
has the form of a single qubit expanded in the Pauli operators, and
represent a physical state when the coefficients satisfy
$\sum\limits_{j=1}^3(\alpha_j^i)^2\le1$. The measurement made by
Alice is the projection on the basis of the \emph{AC} system, which
does not change the matrices $\mu_B^i$  since Eq. (\ref{eq7}) is the
sum of the product $\mu_A^i\otimes\mu_B^i$, and the projection only
affects its probabilities. In any case, Bob's results after the
measurement will be a linear combination of all of $\mu_B^i$.

When the rank of correlation matrix equals unity, this means that
the system \emph{AB} is the direct product state with no
correlations. When the rank of the correlation matrix equals two,
the collapsed state $B^\prime$ after the measurement for any state
\emph{C} is a linear combination of the two $\mu_B^1$ and $\mu_B^2$,
geometrically the state $B^\prime$ is located on the line of
$\mu_B^1$ and $\mu_B^2$, as shown in Fig. 2(a). The line reflects
the one-dimensional nature of the information of the \emph{C}
system. Note the quantum discord (quantum correlation) for rank 2
could be zero. When the rank equals three,the collapsed state
$B^\prime$ after the measurement is linear combination of the three
$\mu_B^1$, $\mu_B^2$ and $\mu_B^3$, geometrically the state
$B^\prime$ is located on the plane of $\mu_B^1$, $\mu_B^2$ and
$\mu_B^3$, as shown in Fig. 2(b), which reflects the two-dimensional
nature of the information contained in the \emph{C}. As we know, the
representation of a pure-state qubit only requires two degrees of
freedom, so rank three is enough to transmit the information of a
pure qubit. When the rank equals four, the collapsed state
$B^\prime$ after the measurement for any state \emph{C} is located
in the three-dimensional tetrahedronal solid space formed by
$\mu_B^1$, $\mu_B^2$, $\mu_B^3$ and $\mu_B^4$, and in three
dimensions the information contained in the \emph{C} can be obtained
from these values. Because an arbitrary single qubit has three
degrees of freedom at most, we can realize the quantum information
transmission of an arbitrary single qubit, when the rank of the
channel system is full. The entanglement of the channel system with
rank four could be zero. That is to say the entanglement is not
necessary for quantum information transmission. Some non-zero
quantum correlation is necessary, as the quantum discord is non-zero
for states of rank 4 and 3 in bipartite systems [15].

\section{An example: Werner state}

Let us now consider a concrete example to implement the quantum
information transmission. If the two-qubit quantum channel, the
\emph{AB} system is in the Werner state [21],
\begin{eqnarray}\label{eq8}
\rho_{AB}=x|\Psi^-\rangle\langle\Psi^-|+\frac{1-x}{4}I_4\quad
x\in[0,1].
\end{eqnarray}
The Werner state is pure and maximally entangled for $x = 1$ with
concurrence (a measure of entanglement) [22] equal to one and
discord also equal to one. As shown analytically in Ref. [23], the
Werner state has both non-zero discord and non-zero concurrence for
$x > 1/3$. In the regime $0 < x \le 1/3$, the Werner state has
non-zero discord, but zero concurrence. Finally for $x = 0$, the
Werner state is a product state with zero discord and zero
concurrence. The correlation matrix of Werner state is $\rm{diag}
[1\quad \emph{x\quad -x\quad x}]$ , which is of full-rank except $x
= 0$. If we do the joint Bell measurement, the state of particle
\emph{B} will collapse to:
\begin{eqnarray}\label{eq9}
\left(
  \begin{array}{cc}
   \frac{1}{2}+(-1)^nx\frac{c_3}{2} & (-1)^mx(\frac{c_1-(-1)^nic_2}{2})  \\
   (-1)^mx(\frac{c_1+(-1)^nic_2}{2}) & \frac{1}{2}-(-1)^nx\frac{c_3}{2}  \\
  \end{array}
\right).
\end{eqnarray}
which contains the information of the qubit $C$. Comparing Eq.
(\ref{eq9}) with Eq. (\ref{eq4}), we have $c_i\propto s_i/x$, the
relation between the received state and the state to be sent. Then
the information of the unknown state \emph{C} can be obtained
directly.

From Eq. (\ref{eq9}), it is easy to see that for $x = 1$ the quantum
information transmission scheme is exactly the traditional quantum
teleportation scheme (except for the unitary operation imposed on
the reduced system B). For $x < 1/3$, the entanglement of the
quantum communication channel is zero, and we can still carry out
the quantum information transmission. In this scheme, we transfer
the complete information needed to reconstruct $\rho_c$, but not the
state itself.

In the information transmission, the output density matrix
$\rho_B^\prime$ is not equal to the input unknown state $\rho_c$,
and cannot be simply transformed into $\rho_c$ via a local unitary
operation as in the ordinary quantum teleportation [1], which is the
price we pay for not having entanglement. Even if $\rho_c$ is a pure
state, the output state ($\rho_B^\prime$) is typically mixed if
$\rho_{AB}$ is not a pure entangled state. In order to know the
received state, Bob must make measurements on the received state,
usually through quantum state tomography [24]. In the ordinary
quantum teleportation, measurement is also needed, if Bob wants to
know the state. Our quantum information transmission plus the local
tomography can be regarded as the remote state \emph{tomography},
which is useful in the quantum information science and technology.

In general, the probabilities of the four Bell measurements are not
equal, which means that Alice can learn something about the state
she is sending, the information could be catch by the eavesdropper
through the classical channel. If the state of the channel has the
following form,
\begin{eqnarray}\label{eq10}
 \rho_{AB}=\frac{1}{2}I_A\otimes\rho_B+\frac{1}{4}\sum\limits_{i,j=1}^3\Lambda_{ij}\sigma_i^A\otimes\sigma_j^B
\end{eqnarray}
where $I_A$ is the identity operator on system \emph{A} and $\rho_B$
is the reduced density matrix of system \emph{B}, the Bell
measurements have equal probability, so that Alice cannot get any
useful information about the sending state from the Bell
measurements. The Werner state can be written in the above form.
This is useful for security consideration.

\section{Conclusions}

In this work we propose a scheme of the quantum information
transmission by using a quantum channel with nonzero quantum
discord. Our scheme becomes ordinary quantum teleportation when a
Bell state is chosen for the quantum channel. In the present scheme,
what transferred from one place to another is the information of the
state with knowing the relation between the received state and the
state to be sent, but not the state itself. This is the disadvantage
of the information transmission. When a particular channel is
chosen, even Alice will have no knowledge on the information she is
sending, while Bob obtains the state with complete information. The
scheme for quantum information transmission can be implemented
experimentally with the Werner states, as the Werner states with
pairs of polarized photons from parametric down-conversion have been
generated [25, 26]. In addition, entangled ions in an ion trap [27]
could also be used for the quantum channel.

\section{Acknowledgments}
We thank Sai Vinjanampathy for helpful discussions and improving the
content of the manuscript. This work is supported by National Basic
Research Program of China (2011CB922203), RGC (HKBU202910), and the
National Natural Science Foundation of China (NSFC) under grant
number (10804042, 10904048). JPD would like to acknowledge support
from the Foundational Questions Institute and the US National
Science Foundation.


%




\end{document}